# Monolayer Magnetic Metal with Scalable Conductivity

*Oleg E. Parfenov,[a] Dmitry V. Averyanov,[a] Ivan S. Sokolov,[a] Alexey N. Mihalyuk,[b,c] Oleg A. Kondratev,[a] Alexander N. Taldenkov,[a] Andrey M. Tokmachev,[a] Vyacheslav G. Storchak [a,*]*

[a] National Research Center "Kurchatov Institute", Kurchatov Sq. 1, 123182 Moscow, Russia
[b] Institute of High Technologies and Advanced Materials, Far Eastern Federal University, 690950 Vladivostok, Russia
[c] Institute of Automation and Control Processes FEB RAS, 690041 Vladivostok, Russia

[*] E-mail: vgstorchak9@gmail.com

2D magnets have emerged as a class of materials highly promising for studies of quantum phenomena and applications in ultra-compact spintronics. Current research aims at design of 2D magnets with particular functional properties. A formidable challenge is to produce metallic monolayers: the material landscape of layered magnetic systems is strongly dominated by insulators; rare metallic magnets, such as $Fe_3GeTe_2$, become insulating as they approach the monolayer limit. Here, electron transport measurements demonstrate that the recently discovered 2D magnet GdAlSi – graphene-like AlSi layers coupled to layers of Gd atoms – remains metallic down to a single monolayer. Band structure analysis indicates the material to be an electride, which may stabilize the metallic state. Remarkably, the sheet conductance of 2D GdAlSi is proportional to the number of monolayers – a manifestation of scalable conductivity. The GdAlSi layers are epitaxially integrated with silicon, facilitating applications in electronics.

# 1. Introduction

The current rise of research on 2D magnetism is expected to bring disruptive technologies in spintronics and optoelectronics.[1-4] Atomically thin intrinsic magnets make the basis for ultra-compact logic and memory devices; they provide a platform for studies of fundamental spin behaviors and unconventional quantum phases.[5,6] A relatively wide range of 2D magnets is now available to researchers, spanning materials with magnetism arising from open *d*-shells[7,8] and *f*-shells[9,10] to extended π-electron systems in nanographenes[11] or moiré structures.[12] An important property of 2D magnets is their high amenability to external stimuli such as magnetic fields,[7] pressure[13] or gating.[8] In view of applications in spintronics, the emphasis is on electron transport: 2D magnets exhibit a number of potentially usable effects such as giant tunneling[14] and lateral[15] magnetoresistance (MR), strong exchange bias.[16,17] Naturally, 2D magnets with various transport properties are required. In the realm of 2D magnets, to be an insulator is the norm; a metallic 2D magnet is a rarity. Yet, this is only part of the picture. Electron transport in 2D magnets depends strongly on the number of monolayers (ML) $N$,[15] especially in the region of a few ML where localization effects become increasingly important. Therefore, it is a formidable challenge to design a monolayer metallic magnet. Moreover, in the 2D limit, it would be advantageous if electron transport were scalable with $N$. Finally, applications require seamless integration of the magnet with technological platforms of modern electronics, such as silicon.

The problem of producing atomically thin metals (ATMs) goes well beyond the field of 2D magnets. It has long been known that the conductivity of bulk metals is typically much higher than that of their thin films or nanowires.[18] For instance, the classical elemental metals, such as Au and Ag, are semiconductors in the ML limit.[19,20] The issue is important for applications in nanoelectronics. The use of the field effect in metals may bring significant advantages over semiconductors but the electric field screening at extremely short distances would require ATM films. In fact, this idea is at the center of the original research on graphene.[21] The field has been further developed in different directions, in particular, to design chemoelectronics employing functionalization of metal nanostructures.[22] Another application where the issue of metal ultra-thin films is critical is the design of interconnects.[23,24] The currently used interconnect lines exhibit a rapidly increasing resistivity at reduced dimensions; their poor performance causes energy consumption and signal delay. The resulting "interconnect bottleneck" is a major limiting factor in the continuous scaling of integrated circuits.[23,24] Therefore, the search for functional ATMs becomes a major challenge in materials science.[18,25,26]

Up-to-date, a number of ATMs have been discovered: for instance, several 2D transition metal dichalcogenides are metallic;[27] metallic nanowires MoP[28] and monolayer MXene flakes[29,30] are promising for nm-thin interconnects. Some of 2D magnets, such as $Fe_3GeTe_2$ (FGT)[8] and 1T-$CrTe_2$,[31] are considered metallic. However, it is necessary to keep in mind the thickness range where the material is a metal as well as the evolution of its electron transport with $N$. A perfect example is FGT, the best known metallic 2D magnet. As $N$ decreases, FGT flakes become increasingly more insulating; in particular, 1 ML of FGT is "very insulating".[8] All in all, these materials do not solve the problem of finding a magnetic ATM with conductivity scalable down to a single ML.

A guiding principle is necessary to design a magnetic ATM. Some recipes are suggested in research on interconnects[18,25,26] but they are rather general to implement. An idea we would like to explore is to employ electrides, materials hosting anionic electrons: it is known that the metallic character of the classical electride $Ca_2N$ is preserved in its 2D flakes.[32] As a blueprint we use the layered system MAlSi which has been predicted to host anionic electrons for M = Ca, Sr, and Ba.[33] The possibility to produce thin films of MAlSi epitaxially integrated with Si has been demonstrated for SrAlSi.[34] To introduce magnetic properties into the system, it is natural to use M with an open shell. We note that rare-earth metalloxenes, $MX_2$ compounds comprising layers of Gd or Eu with those of silicene or germanene, are 2D magnets structurally similar to MAlSi.[9,10] Moreover, the metalloxenes $GdSi_2$ and $EuSi_2$ are metals down to 4 ML.[15] Merging the MAlSi blueprint for electrides with 2D magnetism,[9,10] we suggest layered GdAlSi as a candidate for a magnetic ATM. The stable form of bulk GdAlSi is a non-layered tetragonal polymorph.[35,36] However, polymorph engineering can be employed to produce a layered 2D magnet: it has been shown recently that graphitization results in layered GdAlSi.[37] Moreover, a synthetic route for direct epitaxial integration of layered GdAlSi ($N$ ranging from 11 to 1) with Si is realized; the films exhibit 2D ferromagnetism.[38]

Here, we study the $N$-dependence of electron transport in layered GdAlSi. This 2D magnet is found to be metallic down to a single ML. Remarkably, the sheet conductivity is scalable with $N$ suggesting that the metallic state is robust against localization and interfacial effects.

## 2. Results and Discussion

Our aim is to study electron transport in ultra-thin films of GdAlSi on silicon. In Ref. [38], synthesis of such films and analysis of their structure and magnetic properties have been described in detail. Nevertheless, we believe that it is instructive to provide here a brief account of the synthesis and basic characterization of the samples. Information on GdAlSi magnetism is particularly relevant to discussion of transport properties.

GdAlSi films were produced by molecular beam epitaxy, a technique proved successful in synthesis of lanthanide-based 2D magnets.[9,10,15,17] Synthesis was carried out by direct reaction of the constituent elements: stoichiometric amounts of Gd and Al were deposited on the Si substrate. The substrate was not just a reactant; it also set the orientation of the GdAlSi structure. To synthesize the layered (trigonal) polymorph, we employed the Si(111) face. This choice is important because synthesis on Si(001) results in the tetragonal polymorph of GdAlSi.[39] A series of epitaxial GdAlSi films of a thickness ranging from 1 ML to 11 ML is produced employing the layerwise growth. The upper bound is not quite arbitrary. The trigonal polymorph is stable only below a critical thickness of about 20 ML;[37] as the film becomes thicker, it re-crystallizes into the tetragonal polymorph, the stable form of GdAlSi in the bulk. All the studied films are well below the critical thickness to ensure the absence of the unwanted phase. The quality of the films is probed by diffraction techniques. Figure 1a shows a typical spectrum of reflection high-energy electron diffraction (RHEED). The system of reflexes corresponds to the target product, confirming the epitaxy without any evidence of side phases. The same conclusion follows from X-ray diffraction (XRD) studies (see a typical θ-2θ scan in Figure 1b). Thickness fringes around GdAlSi peaks not only point at superb quality of the films, they provide an independent way to establish the film thickness.

GdAlSi magnetism stems from the triangular lattice of Gd. As a consequence, the magnetic properties of GdAlSi are similar to those of Gd and Eu metalloxenes,[9,10] other materials with the triangular lattice of $4f^7$ ions. In particular, they all exhibit easy-plane magnetic anisotropy. The general trend is that the bulk is an antiferromagnet but the ferromagnetic (FM) signal shows up as the system approaches the 2D limit. The characteristic temperature of the FM transition in 2D GdAlSi is around 40 K. In fact, the effective transition temperature is controlled by weak magnetic fields, as illustrated by Figure 1c. This phenomenon – a fingerprint of 2D ferromagnetism, observed in various 2D magnets[9,10,40] – is particularly strong in the case of easy-plane magnetic anisotropy.[41] The saturation FM moment in 2D GdAlSi (Figure 1d) is well below 7 $\mu_B$/Gd expected for fully spin-polarized Gd ions. The situation is typical for 2D lattices of $4f^7$ ions; it is associated with magnetic phase separation suggested by the exchange bias effect[17] and imaged by magnetic force microscopy.[42]

Having established the atomic structure and magnetism of GdAlSi layers, we focus on their electron transport properties. Here, we studied the film resistivity, the anomalous Hall effect, and MR for various orientations of the external magnetic field – out-of-plane and in-plane, the latter for directions parallel and perpendicular to the current. The results are well reproducible: for instance, the sheet conductivities at 2 K for two separately grown 1 ML films are 235.4 and 228.4 µS/sqr, about twice lower than the values for 2 ML GdAlSi. The most important conclusion is that all the GdAlSi films are metallic. The electron concentration is about $1.5 \cdot 10^{22}$ cm$^2$ V$^{-1}$ s$^{-1}$. They satisfy the Ioffe-Regel criterion for metals ($k_F L > 5$ for all the films). The conductivity of 1 ML GdAlSi is of the same order of magnitude as that of metallic monolayer MXenes[29,30] and metallic TMDC nanosheets.[43-45] Figure 2a demonstrates temperature dependence of resistivity in 1 ML GdAlSi. The dependence is logarithmic; the anisotropy of MR is very weak – these are salient features of a Kondo metal. A very similar pattern is found in 2 ML GdAlSi (Figure S1 of the Supporting Information). However, the situation changes in thicker films: the resistivity maximum at $T_C$, typical of magnetic metals, becomes more pronounced and MR acquires significant anisotropy (Figure S2 of the Supporting Information).

The Kondo metal behavior has been observed in ultra-thin films of magnetic $4f^7$ metalloxenes, GdSi$_2$ and EuSi$_2$, as well.[15] The difference is that the metallic state in those materials does not extend to a single ML: once the film thickness goes below 4 ML, localization effects take center stage. This metal-insulator transition is illustrated in Figure 2b. The resistivity of several ML of GdAlSi, GdSi$_2$ or EuSi$_2$ is rather close but the situation changes radically as the films approach the 1 ML limit. The resistivity of GdAlSi is not very sensitive to the film thickness; in contrast, the resistivity of GdSi$_2$ (or EuSi$_2$) differs drastically between, say, 1 ML and 4 ML films. As a result, the resistivity of 1 ML GdAlSi is 10 orders of magnitude lower than that of 1 ML EuSi$_2$ – a material isoelectronic to GdAlSi. The metal-insulator transition occurs in non-magnetic MSi$_2$ films as well.[46] It seems that the replacement of Si by Al in the anionic layer changes the electronic structure in such a way that the metallic state is stabilized. Therefore, it becomes mandatory to carry out band structure calculations to get insight into the role of Al.

The electronic structure of GdAlSi on Si(111) was assessed employing density functional theory (DFT) calculations (see Methods). The calculations were carried out for $N$ from 1 to 10. All these materials are found to be metals with a significant density of states (DOS) at the Fermi energy. Figure 3a shows the band structure and DOS for 1 ML GdAlSi. The DOS at the Fermi level is dominated by the states in the Gd layer. In the FM phase, the split bands with different spins are well separated in energy (see Figures 3a,b). Analogous data for 2 ML GdAlSi are shown in Figure S3 of the Supporting Information. The number of electrons around the Fermi level depends linearly on $N$ (Figure S4 of the Supporting Information). Post-processing of the calculation results for bulk GdAlSi reveals a significant electron density at the center of the $Gd_3Al_2$ cage (Figures 3c,d), in analogy with other MAlSi compounds.[33] This result suggests a significant density of anionic electrons within the Gd layer, *i.e.* the electride nature of the material. The calculations show the important role of Al: anionic electrons do not show up in the $Gd_3Si_2$ cage (Figure 3d). The same conclusion on the role of Al can be made based on calculations for 1 ML and 2 ML films of GdAlSi (see Figure S5 of the Supporting Information).

1 ML GdAlSi is not just a metal; it is a magnetic metal. Accordingly, its electron transport exhibits specific properties stemming from the magnetic state. In particular, the films demonstrate a significant negative MR. In absolute terms, the MR of 1 ML GdAlSi in out-of-plane magnetic fields decreases as temperature grows (Figure 2c). The longitudinal MR exhibits an analogous trend (Figure S6 of the Supporting Information). The negative MR in thicker films behaves in a similar manner (see Figure S7 of the Supporting Information for MR of 3 ML and 4 ML GdAlSi). At low temperature, MR exhibits a pronounced hysteresis (see Figure 2d for 1 ML and Figure S8 of the Supporting Information for 2 ML GdAlSi).

Another characteristic feature of magnetic materials is the anomalous Hall effect (AHE). The AHE is detected in the $R_{xy}$ measurements for each of the GdAlSi films, from 1 ML to 11 ML. Figure 4a demonstrates evolution of the AHE resistance in 1 ML GdAlSi with temperature. The AHE decreases and vanishes above $T_C$ (see Figure S9 of the Supporting Information). The AHE persists in thicker films as well (see Figure 4b for 9 ML GdAlSi) although the FM moment per Gd decreases as the film becomes thicker.[38] Precise measurements reveal a narrow hysteresis loop (Figure S10 of the Supporting Information).

To assess the evolution and scalability of electron transport, it is necessary to study $N$-dependence in more detail. Figure 5 shows the negative MR as a function of $N$. There is a rapid decrease of the MR (in absolute terms) as the film becomes thicker. This decrease is in agreement with the analogous decline of the FM properties in magnetization measurements.[38] Temperature dependence of the sheet resistance for the films of different $N$ is shown in Figure 6a. All the dependences are very weak, especially in the thicker films. For instance, in 9 ML GdAlSi the changes of resistance below 100 K are within a couple of percent (Figure S2 of the Supporting Information). The changes of the electrical properties with $N$ are best illustrated by Figure 6b demonstrating the $N$-dependence of the sheet conductivity. The dependence is linear with $R^2 \approx 0.99$. Moreover, the y-intersect is an order of magnitude smaller than the slope, *i.e.* the sheet conductance is proportional to N, $\sigma_{xx} = \alpha N$. Thus, the electron transport in GdAlSi is scalable down to a single ML, which is an unusual property. It suggests that electron scattering at surfaces and grain-boundaries is negligible. Search for such materials is a promising research direction in the quest for nanoscale interconnects.[25]

## 3. Conclusion

Research on 2D magnetism is now flourishing; it holds promise for significant advances in fundamentals of magnetism and spintronic applications. To extend the functionality of 2D magnets, it is practical to merge magnetism with other properties; in particular, electron transport is important for electronics. Here, we addressed the challenge of making a magnetic metal with conductivity scalable down to a single ML. We demonstrated that 1 ML GdAlSi is a magnetic metal. The conclusion is based on the results of transport measurements rather than spectroscopic studies commonly used to establish the metallic character of 2D systems.[47,48] The material exhibits scalability of electron transport – the sheet conductance is proportional to N. A possible reason for the unconventional behavior of GdAlSi is that it hosts anionic electrons, according to DFT calculations. Magnetic electrides are suggested for applications in various fields, from spintronics to high-performance catalysis.[49] A significant advantage is that GdAlSi films are seamlessly integrated with Si, the workhorse of modern electronics. It can be important for applications because the metals that do not require any additional layers provide significant conductance benefits.[24] This work may be employed as a blueprint for design of other magnetic ATMs.

## 4. Methods

*Synthesis and Characterization*: The set of GdAlSi films was synthesized in a Riber Compact system for molecular beam epitaxy, employing ultra-high vacuum conditions with the base pressure below $10^{-10}$ Torr. 1 inch × 1 inch Si(111) wafers with miscut angles below 0.5° were used as the substrates. Their temperature was established by thermocouples and a PhotriX ML-AAPX infrared pyrometer. Heating of the substrate at 950 °C removed the natural oxide, as confirmed by the characteristic 7 × 7 surface reconstruction appearing in the RHEED image. GdAlSi layers were synthesized by deposition of Gd and Al on the Si substrate kept at 400 °C. 4N Gd and 5N Al were supplied from Knudsen cell effusion sources heated to 1210 and 905 °C, respectively. The pressures of the reactants in the growth chamber were $1 \cdot 10^{-8}$ Torr for Gd and $6 \cdot 10^{-9}$ Torr for Al, as measured by a Bayard-Alpert ionization gauge. The GdAlSi films were capped by a 20-nm layer of amorphous $SiO_x$, an insulating non-magnetic material that did not affect the electron transport measurements.

Diffraction techniques were used to control the structure of GdAlSi. In the growth chamber, the film surface was monitored by a RHEED diffractometer furnished with the kSA 400 analytical system. After capping, the atomic structure of the films was determined by a Rigaku SmartLab 9 kW XRD diffractometer employing the $CuK_{\alpha 1}$ radiation source. Magnetization measurements of the films were carried out using an MPMS XL-7 SQUID magnetometer. The reciprocating sample option was employed; the diamagnetic contribution of the Si substrate was subtracted in accordance with Ref. [9]. The studies of magnetism and electron transport used macroscopic films of GdAlSi with a lateral size around 5 mm. Electron transport was studied using a Lake Shore 9709A measurement system. Four-contact measurements adhered to the ASTM Standard F76. The ohmicity of the electrical contacts, produced by deposition of an Ag-Ga-Sn alloy, was established by I-V characteristic curves.

*Computational Details*: All calculations in the present study used DFT as implemented in the Vienna *ab initio* simulation package (VASP).[50] The calculations employed the generalized gradient approximation of Perdew, Burke, and Ernzerhof (PBE)[51] as the exchange-correlation functional and the projector-augmented wave approach[52] to describe the electron-ion interaction. In order to determine the ground-state structural model of GdAlSi layers on the Si(111) substrate, we used the *ab initio* random structure searching (AIRSS) method, a straightforward approach to structure prediction based on the stochastic generation of realistic initial structures and their repeated local optimization.[53] To simulate the GdAlSi layers, we employed a silicon slab with the bulk lattice constant optimized by PBE. Hydrogen atoms were used to saturate the dangling bonds at the side of the slab not interfaced with GdAlSi. The kinetic cutoff energy was 300 eV; 12×12×1 *k*-point mesh was used to sample the 1×1 2D Brillouin zone. Geometry optimization was carried out until the residual force on the atoms fell below 10 meV/Å. To improve the description of the partially filled 4*f* states of Gd, we adopted the PBE+U method within Dudarev's formalism[54] with the Hubbard U parameter of 7.0 eV. Two types of Gd pseudopotentials were used: magnetic properties were described employing standard Gd potentials of Ref. [55] whereas the complementary non-magnetic calculations treated 4*f* electrons as core states. Band structure calculations were carried out for in-plane FM ordering of Gd magnetic moments, taking into account spin-orbit coupling. The calculations of the electron density maps demonstrating the anionic electrons used a dense *k*-point mesh of more than 1000 points with a distance between neighboring *k*-points of about 0.03 Å$^{-1}$. The integration range was from the Fermi level down to -2 eV (in the case of bulk GdAlSi) or -0.4 eV (in the case of 1 ML and 2 ML films) with Gaussian smearing of 10 meV.

**Acknowledgements**

This work was supported by NRC "Kurchatov Institute" and the Russian Science Foundation [grants No. 22-13-00004 (synthesis), No. 20-79-10028 (sample characterization), and No. 24-19-00038 (studies of electron transport)]. D.V.A. acknowledges support from the President's scholarship (SP 3111.2022.5). The measurements were carried out using equipment of the resource centers of electrophysical and laboratory X-ray techniques at NRC "Kurchatov Institute". The calculations were conducted using the equipment of Shared Resource Center "Far Eastern Computing Resource" IACP FEB RAS (https://cc.dvo.ru).

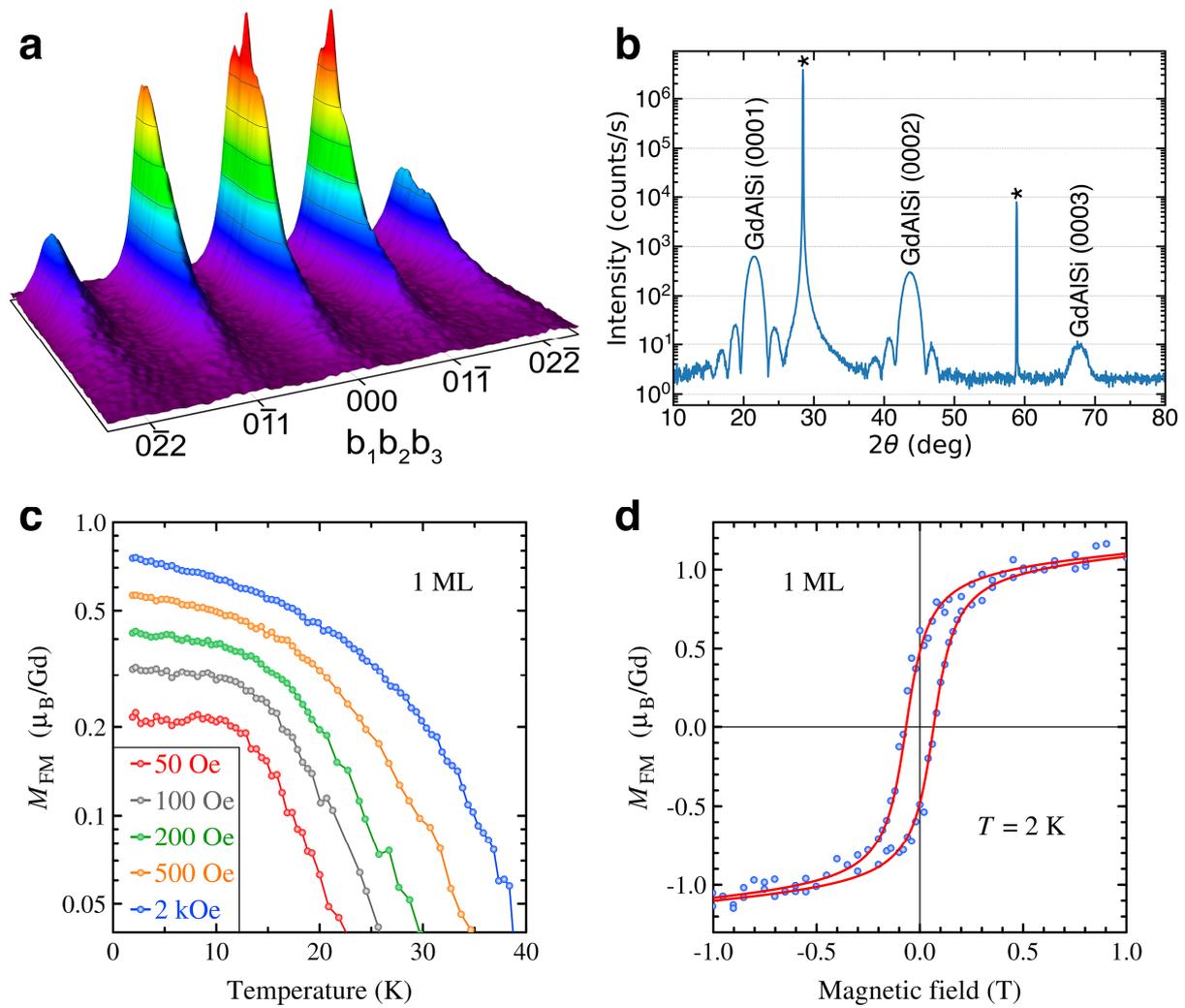

**Figure 1.** a) Typical 3D RHEED image of GdAlSi (1 ML); the reflexes are marked by a triple of Miller-Bravais indices for the basal plane; b) typical θ-2θ XRD scan of GdAlSi (11 ML) demonstrating thickness fringes of the peaks; asterisks mark peaks from the substrate; c) temperature dependence of the FM moment in 1 ML GdAlSi for a selection of magnetic fields: 50 Oe (red), 100 Oe (gray), 200 Oe (green), 500 Oe (orange), and 2 kOe (blue); d) *M-H* hysteresis loop in 1 ML GdAlSi at 2 K.

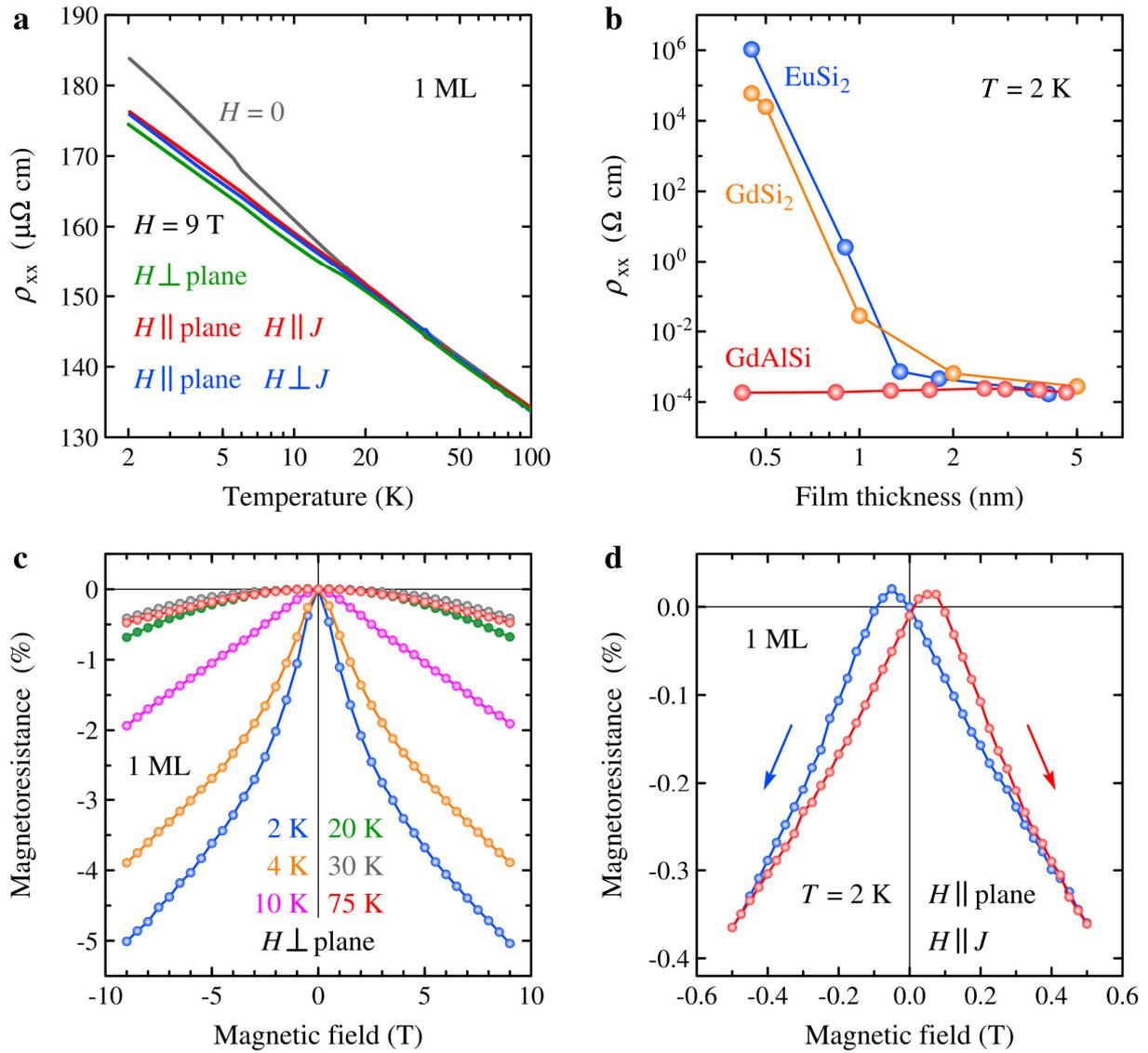

**Figure 2.** a) Temperature dependence of resistivity in 1 ML GdAlSi in zero magnetic field (gray) and magnetic fields of 9 T directed out-of-plane (green), in-plane perpendicular (blue) and parallel (red) to the current $J$; b) dependence of resistivity at 2 K on the film thickness in GdAlSi (red), $GdSi_2$ (orange), and $EuSi_2$ (blue); c) MR of 1 ML GdAlSi in out-of-plane magnetic fields at 2 K (blue), 4 K (orange), 10 K (magenta), 20 K (green), 30 K (gray), and 75 K (red); d) hysteresis in MR of 1 ML GdAlSi at 2 K for in-plane magnetic fields along the current.

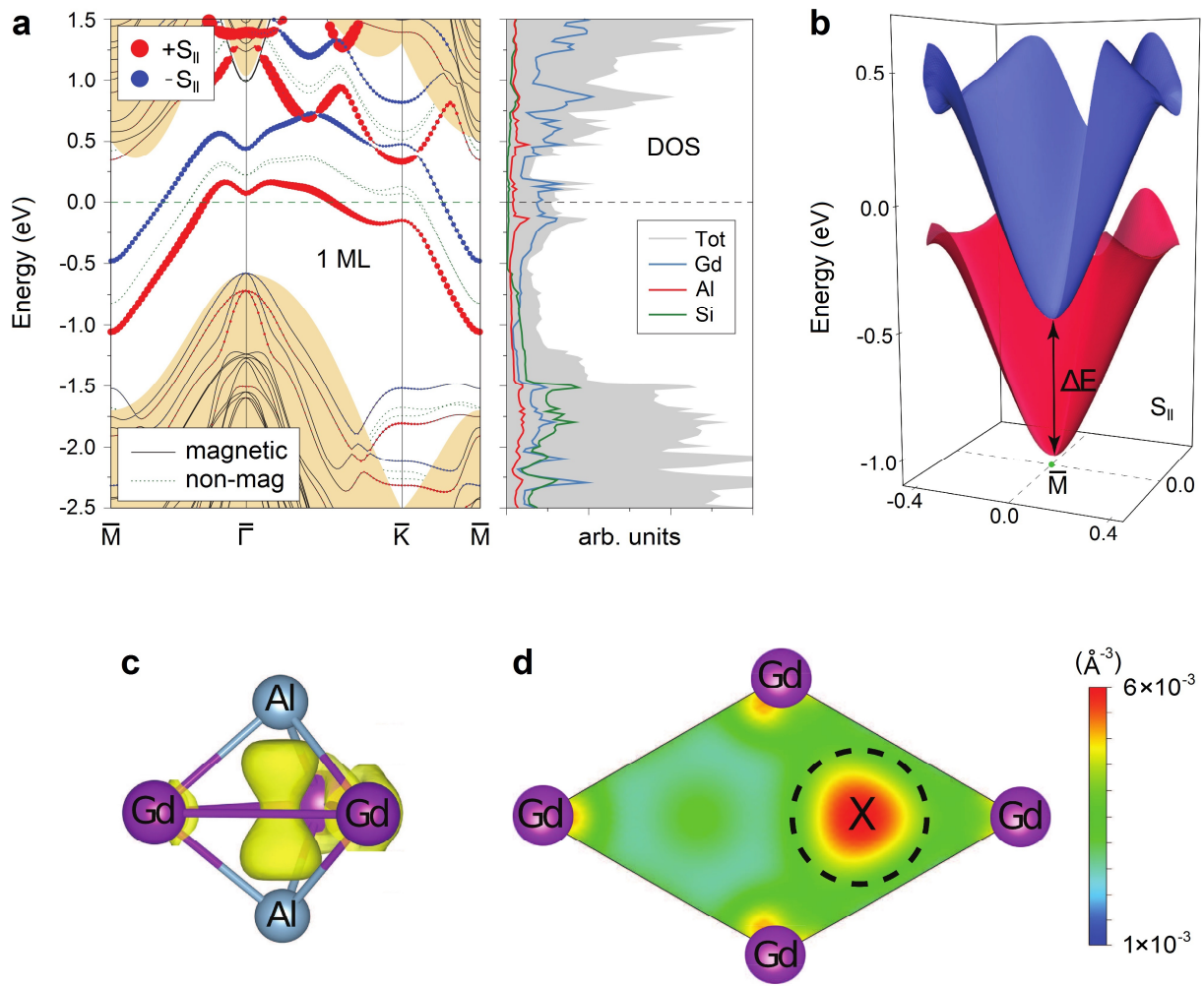

**Figure 3.** a) Band structure of 1 ML GdAlSi and the corresponding element-resolved DOS plot (the Fermi energy is shifted to zero); b) 3D image of the FM split bands of 1 ML GdAlSi around the $\bar{M}$ point in the $k$-space; red and blue colors denote spin-polarized bands with the spins +1/2 and -1/2, respectively; c) electron density of bulk GdAlSi in the $Gd_3Al_2$ cage; d) electron density map of bulk GdAlSi projected onto the (0001) face; the vacant site X, the center of the $Gd_3Al_2$ cage, may host an anionic electron; in the $Gd_3Si_2$ cage (the left part of the rhombus), there is no anionic electron.

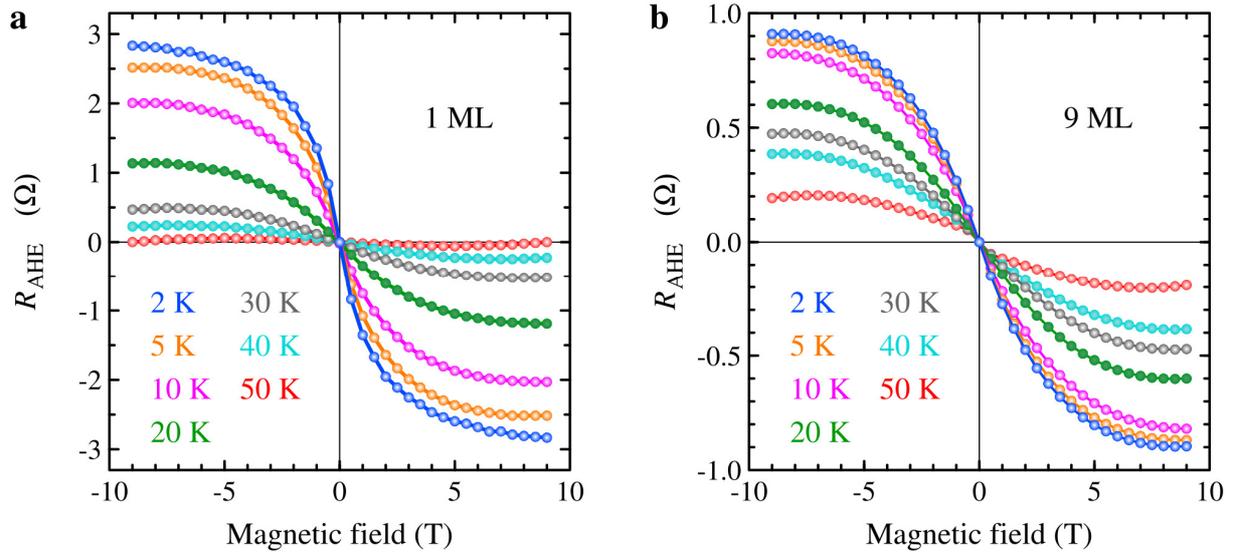

**Figure 4.** Magnetic field dependence of AHE resistance in a) 1 ML and b) 9 ML GdAlSi at 2 K (blue), 5 K (orange), 10 K (magenta), 20 K (green), 30 K (gray), 40 K (cyan), and 50 K (red).

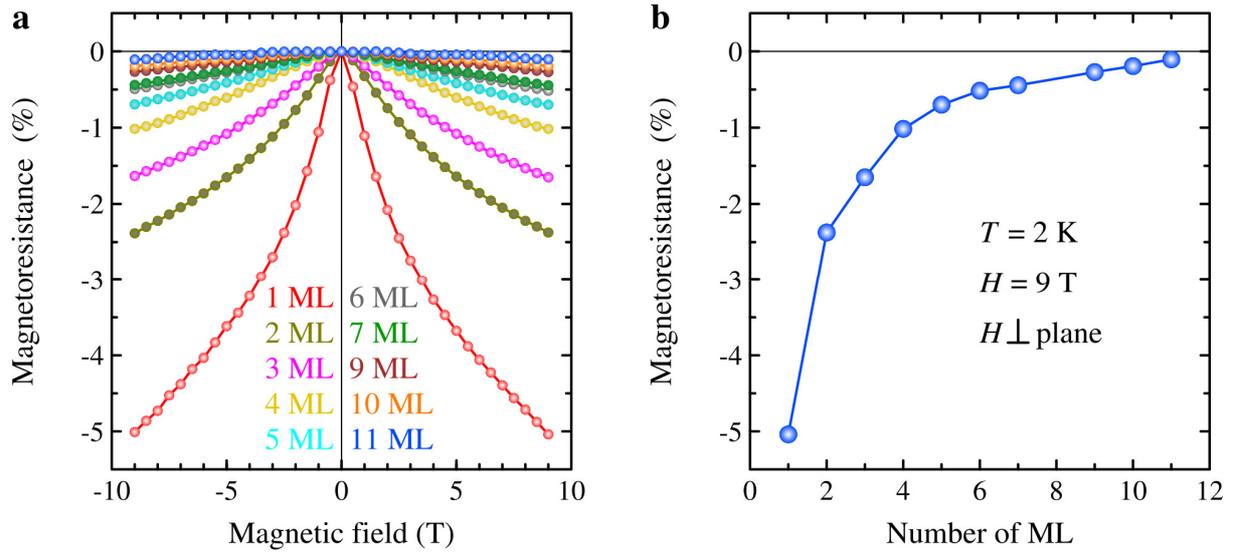

**Figure 5.** a) Dependence of MR at 2 K on out-of-plane magnetic fields in 1 ML (red), 2 ML (olive), 3 ML (magenta), 4 ML (gold), 5 ML (cyan), 6 ML (gray), 7 ML (green), 9 ML (brown), 10 ML (orange), and 11 ML (blue) films of GdAlSi; b) $N$-dependence of MR at 2 K in an out-of-plane magnetic field 9 T.

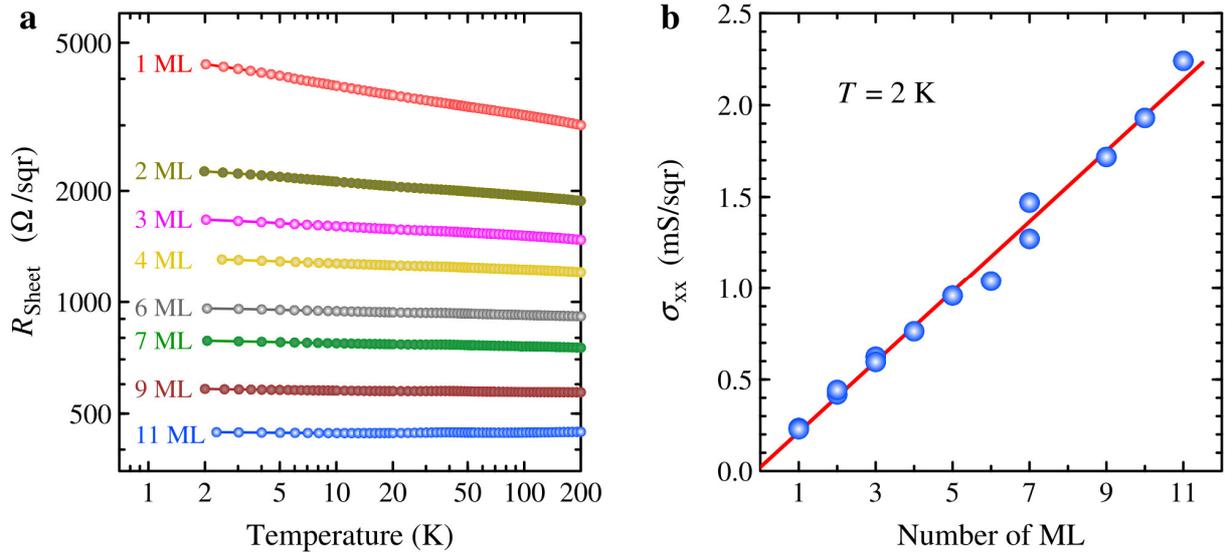

**Figure 6.** a) Temperature dependence of sheet resistance in 1 ML (red), 2 ML (olive), 3 ML (magenta), 4 ML (gold), 6 ML (gray), 7 ML (green), 9 ML (brown), and 11 ML (blue) GdAlSi; b) $N$-dependence of sheet conductance at 2 K in GdAlSi films ranging from 1 ML to 11 ML.

# Supporting Information

**Monolayer Magnetic Metal with Scalable Conductivity**

*Oleg E. Parfenov, Dmitry V. Averyanov, Ivan S. Sokolov, Alexey N. Mihalyuk, Oleg A. Kondratev, Alexander N. Taldenkov, Andrey M. Tokmachev, and Vyacheslav G. Storchak\**

**Content:**

**Figure S1.** Temperature dependence of resistivity in 2 ML GdAlSi.

**Figure S2.** Temperature dependence of resistivity in 9 ML GdAlSi.

**Figure S3.** Band structure of 2 ML GdAlSi.

**Figure S4.** *N*-dependence of the number of electrons at the Fermi level.

**Figure S5.** Maps of electron density in 1 ML and 2 ML GdAlSi.

**Figure S6.** Longitudinal MR of 1 ML GdAlSi.

**Figure S7.** MR of 3 ML and 4 ML GdAlSi.

**Figure S8.** Hysteresis in MR of 2 ML GdAlSi.

**Figure S9.** Temperature dependence of AHE resistance in 1 ML GdAlSi.

**Figure S10.** Hysteresis in AHE resistance of 9 ML GdAlSi.

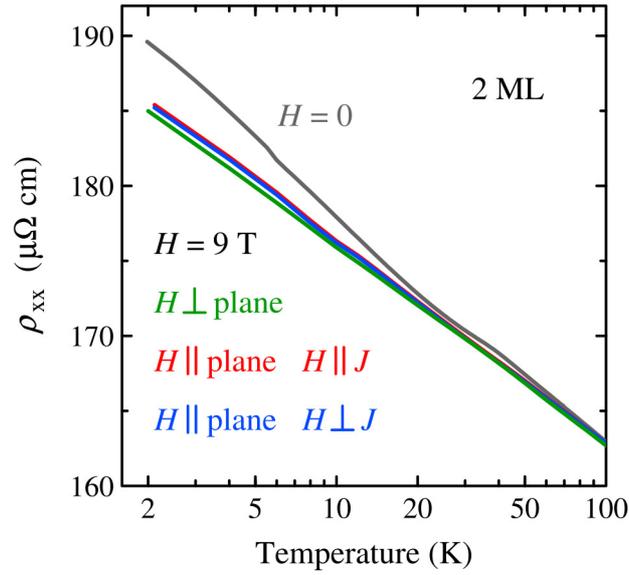

**Figure S1.** Temperature dependence of resistivity in 2 ML GdAlSi in zero magnetic field (gray) and magnetic fields of 9 T directed out-of-plane (green), in-plane perpendicular (blue) and parallel (red) to the current $J$.

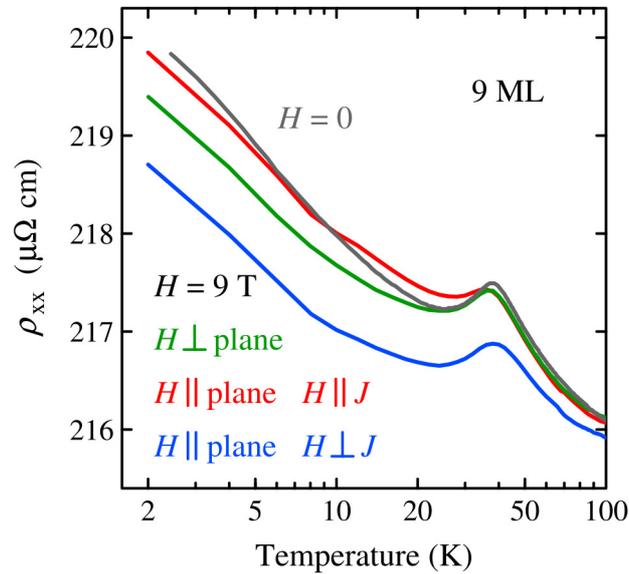

**Figure S2.** Temperature dependence of resistivity in 9 ML GdAlSi in zero magnetic field (gray) and magnetic field of 9 T directed out-of-plane (green), in-plane perpendicular (blue) and parallel (red) to the current $J$.

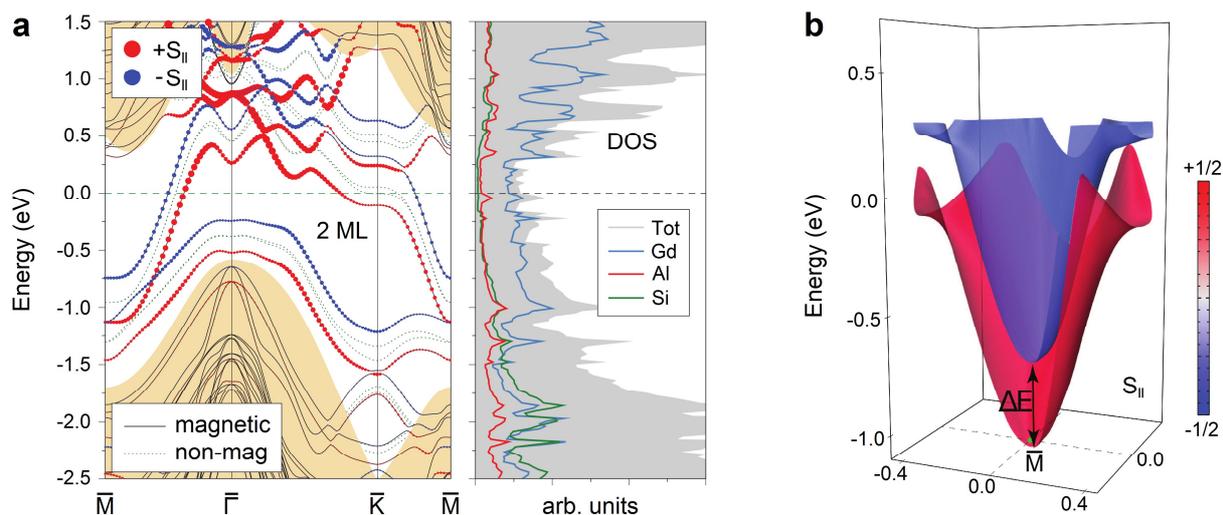

**Figure S3.** a) Band structure of 2 ML GdAlSi and the corresponding element-resolved DOS plot (the Fermi energy is shifted to zero); b) 3D image of the FM split bands of 2 ML GdAlSi around the $\bar{M}$ point in the $k$-space; red and blue colors denote spin-polarized bands with the spins +1/2 and -1/2, respectively.

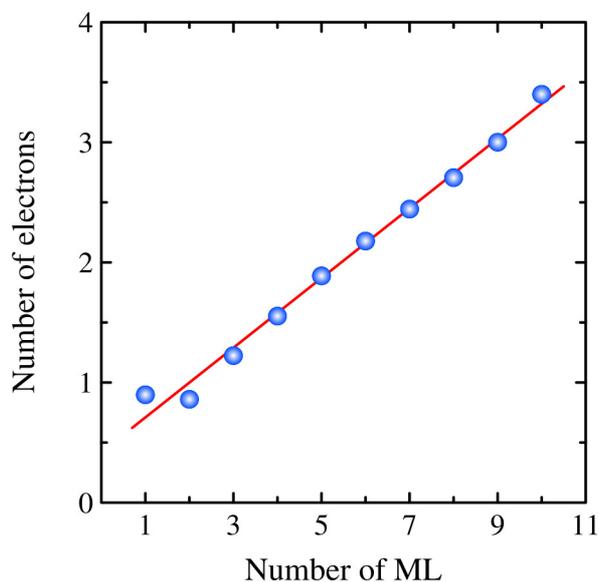

**Figure S4.** $N$-dependence of the number of electrons at the Fermi level (within the range $[E_F - 0.25$ eV$; E_F + 0.25$ eV$]$).

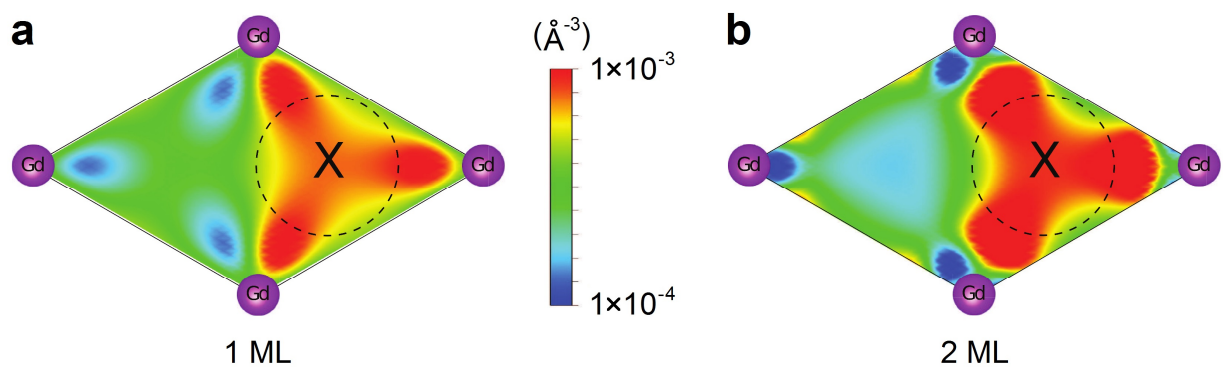

**Figure S5.** Electron density map of a) 1 ML and b) 2 ML GdAlSi projected onto the (0001) face; the vacant site X, the center of the $Gd_3Al_2$ cage, may host an anionic electron; in the $Gd_3Si_2$ cage (the left part of the rhombus), there is no anionic electron.

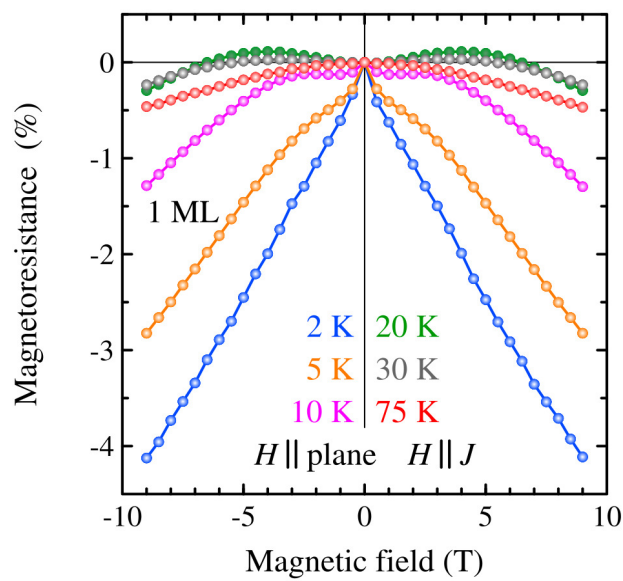

**Figure S6.** Longitudinal MR of 1 ML GdAlSi at 2 K (blue), 5 K (orange), 10 K (magenta), 20 K (green), 30 K (gray), and 75 K (red).

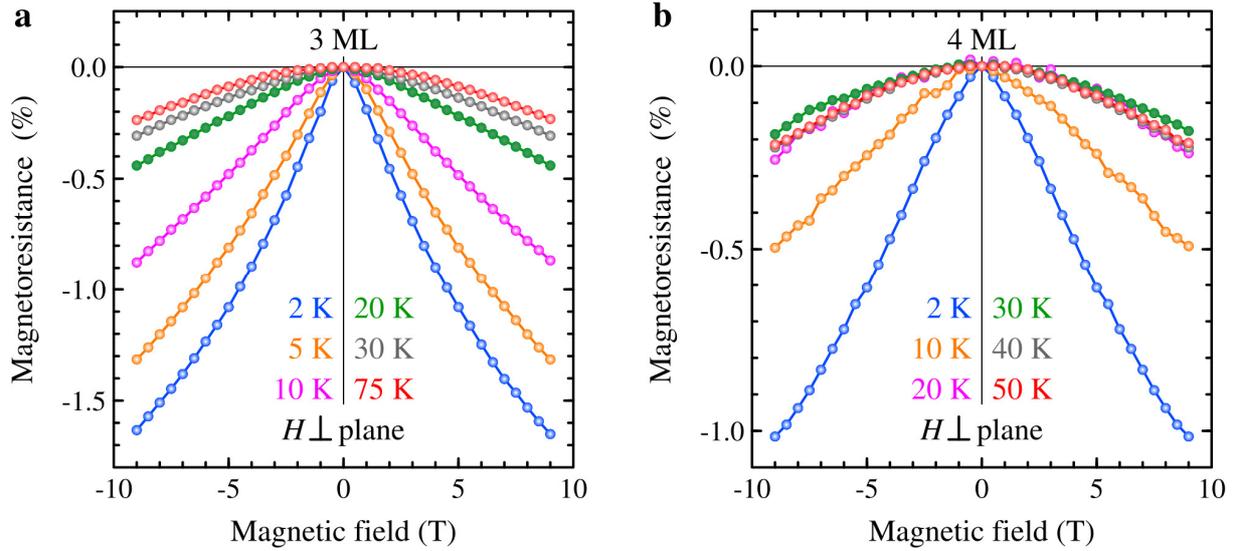

**Figure S7.** a) MR of 3 ML GdAlSi in out-of-plane magnetic fields at 2 K (blue), 5 K (orange), 10 K (magenta), 20 K (green), 30 K (gray), and 75 K (red); b) MR of 4 ML GdAlSi in out-of-plane magnetic fields at 2 K (blue), 10 K (orange), 20 K (magenta), 30 K (green), 40 K (gray), and 50 K (red).

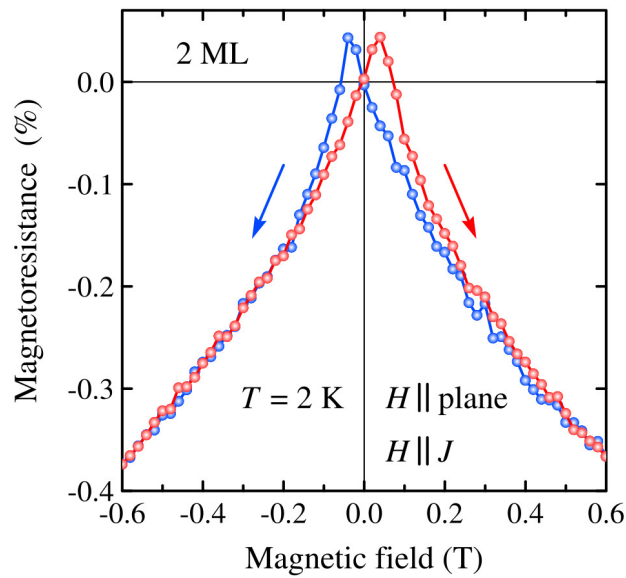

**Figure S8.** Hysteresis in longitudinal MR of 2 ML GdAlSi at 2 K.

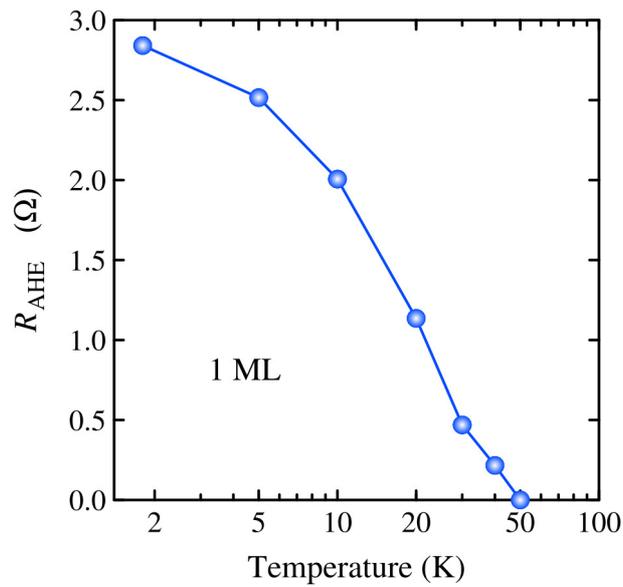

**Figure S9.** Temperature dependence of AHE resistance at -9 T in 1 ML GdAlSi.

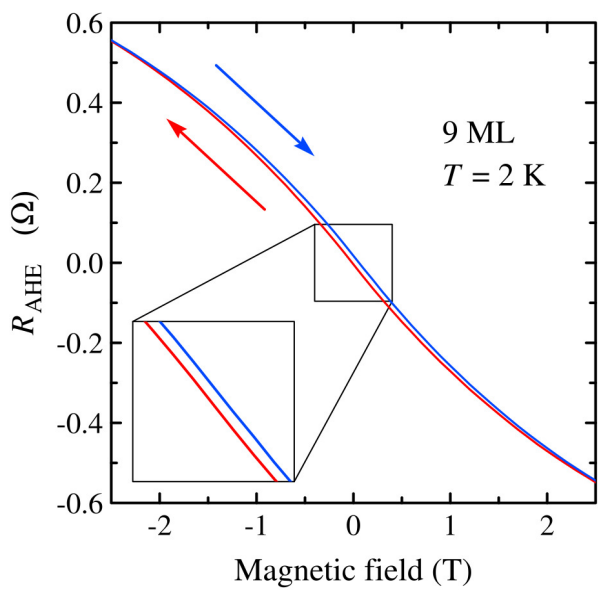

**Figure S10.** Hysteresis in AHE resistance of 9 ML GdAlSi at 2 K.